\documentstyle[sprocl]{article}

\def\Journal#1#2#3#4{{#1} {\bf #2}, #3 (#4)}

\def\PRD{{\em Phys. Rev.} D}

\def\be{\begin{equation}}
\def\ee{\end{equation}}
\def\bea{\begin{eqnarray}}
\def\eea{\end{eqnarray}}
\def\lab#1      {\hbox{\small #1} }

\begin{document}

\title{Vortex configurations in $SO(3) \times Z(2)$}

\author{Andrei Alexandru\footnote{Talk presented
at Confinement IV, Vienna, July 3 - 8 by A. Alexandru.
Supported in part by U.S. Dept. of Energy grant DE-FG05-91 ER 40617.} and Richard W. Haymaker}

\address{Department of Physiscs and Astronomy,\\Louisiana State University, Baton Rouge, Louisisana 70803-4001, USA \\
alexan@rouge.phys.lsu.edu, haymaker@rouge.phys.lsu.edu}


\maketitle\abstracts{We study the configuration space of the Tomboulis $SO(3) \times Z(2)$ formulation
with periodic boundary conditions. The dynamical variables are constrained by the 
required coincidence of Z(2) and SO(3) monopoles. Furthermore, there is an additional constraint coming from the boundary conditions. We propose an update algorithm 
that satisfies the constraints and is straightforward to implement.}

\section{Introduction}
Center vortices (i.e. configurations formed out of the center of the gauge group) are a disordering mechanism 
which can lead to an area law for Wilson loops.  We examine here the gauge invariant description 
of these vortices derived by Tomboulis and others[1-3].   In particular 
we explore the configuration space in this formulation and present an algorithm for simulations.

The basic idea behind the Tomboulis formulation is to write an $SU(N)$ gauge theory in terms of $SU(N)/Z(N)$ and $Z(N)$ variables separately. The $SU(N)/Z(N)$ variables live on links and the $Z(N)$ variables live on plaquettes. 
This results in an expanded set of variables in the partition function but they are related by constraints.
Tomboulis determined the constraints for a lattice with free boundary conditions. However, in numerical simulations we use periodic boundary conditions and thus it is important to see how this new constraints limit the 
configuration space[4].

\section{Simplified version of Tomboulis derivation}
We restrict our attention here to the $SU(2)$ gauge group.  
We will present here a simpler derivation than the ones in Refs.[2,4].

We start with the usual partition function:
$$
Z=\int [dU]_b e^{\frac{\beta}{2}\sum_p Tr(U_p)}
$$
where $[dU]_b=\prod_b dU_b$ is the Haar invariant measure and $U_p$ is the usual plaquette definition and the trace is taken in the fundamental representation of $SU(2)$. Now $[dU]_b$ is invariant under the transformation $U_b\rightarrow \gamma_b U_b$ where $\gamma_b\in Z(2)$. We then obtain:
$$
Z=\int [dU]_b e^{\frac{\beta}{2}\sum_p Tr(U_p) \bar{\gamma}(p)}
$$
where $\bar{\gamma}(p)=\prod_{b\in \partial p} \gamma_b$.

This is true for any $\gamma_b$. Each $\gamma_b$ generates a $\bar{\gamma}(p)$. We see that where $\gamma$ is defined on links $\bar\gamma$ is defined on plaquettes. The entire set of $\gamma$ defined on links will induce a subset of configurations defined on plaquettes. We will denote this set with ${\cal D}$. For any $\sigma\in{\cal D}$ we have:
$$
Z=\int [dU]_b e^{\frac{\beta}{2}\sum_p Tr(U_p)\sigma(p)}
$$
Since this is true for any $\sigma\in{\cal D}$ the we can write:
$$
Z=\frac{1}{|{\cal D}|}\int [dU]_b \sum_{\sigma\in{\cal D}} e^{\frac{\beta}{2}\sum_p Tr(U_p) \sigma(p)}
$$
where $|{\cal D}|$ is the number of elements in ${\cal D}$.

One last step is to define $\eta(p)= \lab{sgn} (Tr(U_p))$. Now write $\sigma'=\sigma\eta$. We will have:
$$
Z=\frac{1}{|{\cal D}|}\int [dU]_b \sum_{\sigma'\eta\in{\cal D}} e^{\frac{\beta}{2}\sum_p |Tr(U_p)| \sigma'(p)}
$$
If we look at the final form of the action we see that it is invariant under $U_b\rightarrow -U_b$. 
%Thus instead of integrating over all $SU(2)$ we can choose to integrate only a half of it, 
%i.e. integrate over $SU(2)/Z(2) = SO(3)$. 
Summing everything up we get and droping the prime in $\sigma'$ we get:
\begin{equation}
Z=\frac{2^{N_b}}{|{\cal D}|}\int [dU]'_b \sum_{\sigma\eta\in{\cal D}} e^{\frac{\beta}{2}\sum_p |Tr(U_p)| \sigma(p)} \label{eq1}
\end{equation}
where $N_b$ is the number of links and $[dU]'_b$ denotes integration over $SO(3)$.

We see that now we have a theory that is formulated in terms of $SO(3)$ variables living on links and $Z_2$ variables living on plaquettes. The $SO(3)$ variables are unconstrained and the $Z(2)$ variables are constrained by the set ${\cal D}$. This set describes the constraints that the vortices have to obey.

\section{Properties of ${\cal D}$}
In this derivation we see that ${\cal D}$ is simply the space generated by link flips i.e. the allowed configurations space can be defined constructively by saying that an allowed configuration has to be the product of succesive star transformations. A star transformation around a link b is a transformation that changes the sign simultaneously on all plaquettes that form the co-boundary of b. This is a useful definition since it provides us with a method to simulate this numerically.

Since all allowed configurations are constructed using star transformations it implies that all allowed configurations obey what we call the cubic constraint, i.e. the product of plaquettes that form the faces of any cube has to be unity. This is equivalent with saying that the thin monopoles (i.e. cubes with $\sigma(\partial c)=-1$) have to coincide with the thick monopoles (i.e. cubes with $\eta(\partial c)=-1$).

One immediate question is whether a configuration that obeys cubic constraint is automatically allowed. Tomboulis [2] showed that this is true for the free boundary conditions. However in the case of periodic (antiperiodic) boundary conditions we proved that this is no longer true [4]. Basically we showed that for a configuration to be allowed in the case of periodic boundary conditions the vortices have to wrap an even number of times around the lattice.  

\section{Vortex counters in Tomboulis formulation} 
If we look at the partition function (\ref{eq1}) we see that there are two types of $Z(2)$ variables: $\sigma$ and $\eta$. Tomboulis identifies the $\sigma$ structures with thin vortices. 

The thick vortices are defined using the Wilson loop. In Tomboulis formulation the Wilson loop has some extra factors:
$$
\langle W(C)\rangle=\langle\frac{1}{2}(\prod_{b\in C}U_b) \sigma(S)\eta(S)\rangle
$$
where $S$ is any surface with the boundary $C$. We say that the Wilson loop entraps an odd number of thick vortices if $\lab{sgn} \left[\prod_{b\in C}U_b \right] \eta(S)=-1$ for any surface $S$ with $\partial S=C$.

Besides the usual thick and thin vortices we have the hybrid vortex, formed from patches of each type.

\end{document}